# Coordinate, Momentum and Dispersion operators in Phase space representation

**[1]Hanitriarivo Rakotoson**, **[2]Raoelina Andriambololona**, **[3]RavoTokiniaina Ranaivoson**,**[4]Raboanary Roland**

[1]infotsara@gmail.com;[2]raoelinasp@yahoo.fr; [2]jacquelineraoelina@hotmail.com; [2]raoelina.andriambololona@gmail.com
[3]tokhiniaina@gmail.com;[3]tokiniainaravor13@gmail.com; [4]r_raboanary@yahoo.fr

[1,2,3]Theoretical Physics Department, Institut National des Sciences et Techniques Nucléaires (INSTN- Madagascar), BP 4279 101, Antananarivo –Madagascar,

[4]Department of Physics, University of Antananarivo-Madagascar.

---

***Abstract-The aim of this paper is to present a study on the representations of coordinate, momentum and dispersion operators in the framework of a phase space representation of quantum mechanics that we have introduced and studied in previous works. We begin in the introduction section with a recall about the concept of representation of operators on wave function spaces. Then, we show that in the case of the phase space representation the coordinate and momentum operators can be represented either with differential operators or with matrices. The explicit expressions of both the differential operators and matrices representations are established. Multidimensional generalization of the obtained results are performed and phase space representation of dispersion operators are given.***

***Keywords :*** quantum theory, phase space, quantum operators, representation, matrices

---

## I. INTRODUCTION

In our previous work [1] we have performed a study on a phase space representation of quantum theory. For the case of one dimensional quantum mechanics we have introduced a basis denoted $\{\,|n,X,P,\mathscr{b}\rangle\}$ of the state space to define this phase space representation [1], [2]. A vector basis $|n,X,P,\mathscr{b}\rangle$ designates an eigenstate of operators called coordinate and momentum dispersion operators denoted $\mathbf{\Sigma}_x$ and $\mathbf{\Sigma}_p$:

$$\begin{cases}\mathbf{\Sigma}_x = \dfrac{1}{2}[\dfrac{(\boldsymbol{x}-X)^2}{(a)^2}+\dfrac{(\boldsymbol{p}-P)^2}{(\mathscr{b})^2}](a)^2 \\ \mathbf{\Sigma}_p = \dfrac{1}{2}[\dfrac{(\boldsymbol{x}-X)^2}{(a)^2}+\dfrac{(\boldsymbol{p}-P)^2}{(\mathscr{b})^2}](\mathscr{b})^2\end{cases} \quad (1.1)$$

$$\begin{cases}\mathbf{\Sigma}_x|n,X,P,\mathscr{b}\rangle = (2n+1)(a)^2|n,X,P,\mathscr{b}\rangle \\ \mathbf{\Sigma}_p|n,X,P,\mathscr{b}\rangle = (2n+1)(\mathscr{b})^2|n,X,P,\mathscr{b}\rangle\end{cases} \quad (1.2)$$

$$a\mathscr{b} = \frac{\hbar}{2} \quad (1.3)$$

In these expressions, $\boldsymbol{x}$ and $\boldsymbol{p}$ are the coordinate and momentum operators [1],[2],[3]. $X$ and $P$ are respectively the mean values of $\boldsymbol{x}$ and $\boldsymbol{p}$ in a state $|n,X,P,\mathscr{b}\rangle$

$$\begin{cases}X = \langle n,X,P,\mathscr{b}|\boldsymbol{x}|n,X,P,\mathscr{b}\rangle \\ P = \langle n,X,P,\mathscr{b}|\boldsymbol{p}|n,X,P,\mathscr{b}\rangle\end{cases} \quad (1.4)$$

and the eigen values $(2n+1)(a)^2$ and $(2n+1)(\mathscr{b})^2$ of $\mathbf{\Sigma}_x$ and $\mathbf{\Sigma}_p$ are the coordinate and momentum statistical variance (statistical dispersion) for a state $|n,X,P,\mathscr{b}\rangle$

$$\begin{cases}(2n+1)(a)^2 = \langle n,X,P,\mathscr{b}|(\boldsymbol{x}-X)^2|n,X,P,\mathscr{b}\rangle \\ (2n+1)(\mathscr{b})^2 = \langle n,X,P,\mathscr{b}|(\boldsymbol{p}-P)^2|n,X,P,\mathscr{b}\rangle\end{cases} \quad (1.5)$$

Let us denote $|x\rangle$ an eigenstate of the coordinate operator $\boldsymbol{x}$ and $|p\rangle$ an eigenstate of the momentum operator $\boldsymbol{p}$. The basis $\{|x\rangle\}$ and $\{|p\rangle\}$ define respectively the coordinate and momentum representations.

Let $\mathcal{E}$ be the state space of the particle. For any element $|\psi\rangle$ of $\mathcal{E}$ we have the decomposition [1],[4],[5],[6]

$$|\psi\rangle = \int |x\rangle\langle x|\psi\rangle\, dx = \int \psi(x)\,|x\rangle dx \quad (1.6)$$

$$= \int |p\rangle\langle p|\psi\rangle\, dp = \int \tilde{\psi}(p)\,|p\rangle dp \quad (1.7)$$

$\psi(x) = \langle x|\psi\rangle$ and $\tilde{\psi}(p) = \langle p|\psi\rangle$ are the wave functions corresponding to the state $|\psi\rangle$ respectively in coordinate and momentum representation. We will denote $\hat{\mathcal{E}}$ and $\tilde{\mathcal{E}}$ the wave function spaces i.e the sets of wave functions respectively in coordinate and momentum representation $\hat{\mathcal{E}} = \{\psi\}$ and $\tilde{\mathcal{E}} = \{\tilde{\psi}\}$.

Let $\boldsymbol{A}$ be a quantum observable of the particle i.e a hermitian linear operator on the space state $\mathcal{E}$[3],[7],[8],[9]. We may write

$$\boldsymbol{A}|\psi\rangle = \int |x\rangle\langle x|\boldsymbol{A}|\psi\rangle\, dx = \int [\widehat{\boldsymbol{A}}\psi(x)]|x\rangle\, dx \quad (1.8)$$

$$= \int |p\rangle\langle p|\boldsymbol{A}|\psi\rangle\, dp = \int [\widetilde{\boldsymbol{A}}\tilde{\psi}(p)]|p\rangle\, dp \quad (1.9)$$

The relations defining the concept of representations of the operator $\boldsymbol{A}$ on wave functions spaces are

$$\langle x|\boldsymbol{A}|\psi\rangle = \widehat{\boldsymbol{A}}\psi(x) \quad (1.10)$$

$$\langle p|\boldsymbol{A}|\psi\rangle = \widetilde{\boldsymbol{A}}\tilde{\psi}(p) \quad (1.11)$$

In these expressions, $\widehat{\boldsymbol{A}}$ is a linear operator over the wave functions space $\hat{\mathcal{E}}$ (coordinate representation) and $\widetilde{\boldsymbol{A}}$ a linear operator over the wave functions space $\tilde{\mathcal{E}}$ (momentum representation). So $\widehat{\boldsymbol{A}}$ is the representation of $\boldsymbol{A}$ over $\hat{\mathcal{E}}$ and $\widetilde{\boldsymbol{A}}$ its representation over $\tilde{\mathcal{E}}$. For the case of the coordinate and momentum operator $\boldsymbol{x}$ and $\boldsymbol{p}$ themselves, we have

$$\boldsymbol{x}|\psi\rangle = \int |x\rangle\langle x|\boldsymbol{x}|\psi\rangle\, dx = \int [x\psi(x)]\, |x\rangle dx \tag{1.12}$$

$$= \int |p\rangle\langle p|\boldsymbol{x}|\psi\rangle\, dp = \int [i\hbar \frac{\partial\tilde{\psi}(p)}{\partial p}]\, |p\rangle dp \tag{1.13}$$

$$\boldsymbol{p}|\psi\rangle = \int |x\rangle\langle x|\boldsymbol{p}|\psi\rangle\, dx = \int [-i\hbar \frac{\partial\psi(x)}{\partial x}]|x\rangle\, dx \tag{1.14}$$

$$= \int |p\rangle\langle p|\boldsymbol{p}|\psi\rangle\, dp = \int [p\tilde{\psi}(p)]\, |p\rangle dp \tag{1.15}$$

so

$$\begin{cases} \widehat{\boldsymbol{x}} = x \\ \widehat{\boldsymbol{p}} = -i\hbar\dfrac{\partial}{\partial x} \end{cases} \qquad \begin{cases} \widetilde{\boldsymbol{x}} = i\hbar\dfrac{\partial}{\partial p} \\ \widetilde{\boldsymbol{p}} = p \end{cases} \tag{1.16}$$

we have for all the operator couples $(\boldsymbol{x},\boldsymbol{p})$, $(\widehat{\boldsymbol{x}},\widehat{\boldsymbol{p}})$ and $(\widetilde{\boldsymbol{x}},\widetilde{\boldsymbol{p}})$ [3],[10],[11],[12]

$$\begin{cases} [\boldsymbol{x},\boldsymbol{p}]_- = i\hbar \\ [\widehat{\boldsymbol{x}},\widehat{\boldsymbol{p}}]_- = i\hbar \\ [\widetilde{\boldsymbol{x}},\widetilde{\boldsymbol{p}}]_- = i\hbar \end{cases} \tag{1.17}$$

If we consider now the case of the phase space representation defined with the basis $\{|n,X,P,\mathscr{b}\rangle\}$, following our previous works [1],[2], we have for any state $|\psi\rangle$ of the particle and for any fixed value of $n$

$$|\psi\rangle = \int |n,X,P,\mathscr{b}\rangle\langle n,X,P,\mathscr{b}|\psi\rangle \frac{dXdP}{2\pi\hbar}$$

$$= \int |n,X,P,\mathscr{b}\rangle\Psi^n(X,P,\mathscr{b}) \frac{dXdP}{2\pi\hbar} \tag{1.18}$$

$\Psi^n(X,P,\mathscr{b}) = \langle n,X,P,\mathscr{b}|\psi\rangle$ being the wave function corresponding to the state $|\psi\rangle$. Let us denote $\overset{\frown}{\tilde{\mathcal{E}}}$ the set of the $\Psi^n$. For any quantum observable $\boldsymbol{A}$ of the particle, we may introduce the representation $\overset{\frown}{\tilde{\boldsymbol{A}}}$ of $\boldsymbol{A}$ over $\overset{\frown}{\tilde{\mathcal{E}}}$ such as

$$\boldsymbol{A}|\psi\rangle = \int |n,X,P,\mathscr{b}\rangle\langle n,X,P,\mathscr{b}|\boldsymbol{A}|\psi\rangle \frac{dXdP}{2\pi\hbar}$$

$$= \int [\overset{\frown}{\tilde{\boldsymbol{A}}}\Psi^n(X,P,\mathscr{b})]|n,X,P,\mathscr{b}\rangle \frac{dXdP}{2\pi\hbar} \tag{1.19}$$

$$\langle n,X,P,\mathscr{b}|\boldsymbol{A}|\psi\rangle = \overset{\frown}{\tilde{\boldsymbol{A}}}\Psi^n(X,P,\mathscr{b}) \tag{1.20}$$

In our work [1], we have also established that another possible expression for the decomposition of the state $|\psi\rangle$ in the phase space representation is

$$|\psi\rangle = \sum_n |n,X,P,\mathscr{b}\rangle\langle n,X,P,\mathscr{b}|\psi\rangle$$

$$= \sum_n |n,X,P,\mathscr{b}\rangle\Psi^n(X,P,\mathscr{b}) \tag{1.21}$$

So we may also define a matrix representation $[\boldsymbol{A}]$ of an observable $\boldsymbol{A}$ in the basis $\{|n,X,P,\mathscr{b}\rangle\}$,

$$\boldsymbol{A} = \sum_n\sum_m A_{\boldsymbol{m}}^{\boldsymbol{n}}|n,X,P,\mathscr{b}\rangle\,\langle m,X,P,\mathscr{b}| \tag{1.22}$$

$$\boldsymbol{A}|\psi\rangle = \sum_n\sum_m A_{\boldsymbol{m}}^{\boldsymbol{n}}\Psi^m(X,P,\mathscr{b})\,|n,X,P,\mathscr{b}\rangle \tag{1.23}$$

with $A_{\boldsymbol{m}}^{\boldsymbol{n}}$ the elements of the matrix $[\boldsymbol{A}]$

$$A_{\boldsymbol{m}}^{\boldsymbol{n}} = \langle n,X,P,\mathscr{b}|\boldsymbol{A}|m,X,P,\mathscr{b}\rangle \tag{1.24}$$

$$\boldsymbol{A}|m,X,P,\mathscr{b}\rangle = \sum_n A_{\boldsymbol{m}}^{\boldsymbol{n}}\,|n,X,P,\mathscr{b}\rangle \tag{1.25}$$

## II. DIFFERENTIAL OPERATOR REPRESENTATION OF THE COORDINATE AND MOMENTUM OPERATORS

Let $\overset{\frown}{\tilde{\boldsymbol{x}}}$ and $\overset{\frown}{\tilde{\boldsymbol{p}}}$ be the representations of the coordinate and momentum operators $\boldsymbol{x}$ and $\boldsymbol{p}$ over the phase space wave functions space $\overset{\frown}{\tilde{\mathcal{E}}}$

$$\langle n,X,P,\mathscr{b}|\boldsymbol{x}|\psi\rangle = \overset{\frown}{\tilde{\boldsymbol{x}}}\,\Psi^n(X,P,\mathscr{b}) \tag{2.1}$$

$$\langle n,X,P,\mathscr{b}|\boldsymbol{p}|\psi\rangle = \overset{\frown}{\tilde{\boldsymbol{p}}}\,\Psi^n(X,P,\mathscr{b}) \tag{2.2}$$

As in our work [2] we may also introduce the operators $\boldsymbol{\mathcal{x}}$ and $\boldsymbol{\mathcal{p}}$ defined by the relations

$$\begin{cases} \boldsymbol{\mathcal{x}} = \dfrac{\boldsymbol{x} - X}{\sqrt{2}a} \\ \boldsymbol{\mathcal{p}} = \dfrac{\boldsymbol{p} - P}{\sqrt{2}\mathscr{b}} \end{cases} \tag{2.3}$$

and their representation $\overset{\frown}{\tilde{\boldsymbol{\mathcal{x}}}}$ and $\overset{\frown}{\tilde{\boldsymbol{\mathcal{p}}}}$ over $\overset{\frown}{\tilde{\mathcal{E}}}$

$$\langle n,X,P,\mathscr{b}|\boldsymbol{\mathcal{x}}|\psi\rangle = \overset{\frown}{\tilde{\boldsymbol{\mathcal{x}}}}\,\Psi^n(X,P,\mathscr{b}) \tag{2.4}$$

$$\langle n,X,P,\mathscr{b}|\boldsymbol{\mathcal{p}}|\psi\rangle = \overset{\frown}{\tilde{\boldsymbol{\mathcal{p}}}}\,\Psi^n(X,P,\mathscr{b}) \tag{2.5}$$

$$\begin{cases} \overset{\frown}{\tilde{\boldsymbol{\mathcal{p}}}} = \dfrac{\overset{\frown}{\tilde{\boldsymbol{p}}} - P}{\sqrt{2}\mathscr{b}} \\ \overset{\frown}{\tilde{\boldsymbol{\mathcal{x}}}} = \dfrac{\overset{\frown}{\tilde{\boldsymbol{x}}} - X}{\sqrt{2}a} \end{cases} \tag{2.6}$$

In the work [2],we establish that for the operators

$$\begin{cases} \boldsymbol{\mathcal{z}}^- = \dfrac{1}{\sqrt{2}}[\boldsymbol{\mathcal{p}} - i\boldsymbol{\mathcal{x}}] \\ \boldsymbol{\mathcal{z}}^+ = \dfrac{1}{\sqrt{2}}[\boldsymbol{\mathcal{p}} + i\boldsymbol{\mathcal{x}}] \end{cases} \Leftrightarrow \begin{cases} \boldsymbol{\mathcal{p}} = \dfrac{1}{\sqrt{2}}[\boldsymbol{\mathcal{z}}^- + \boldsymbol{\mathcal{z}}^+] \\ \boldsymbol{\mathcal{x}} = \dfrac{i}{\sqrt{2}}[\boldsymbol{\mathcal{z}}^- - \boldsymbol{\mathcal{z}}^+] \end{cases} \tag{2.7}$$

we have

$$\begin{cases} \boldsymbol{\mathcal{z}}^-|n,X,P,\mathscr{b}\rangle = \sqrt{n}|n-1,X,P,\mathscr{b}\rangle \\ \boldsymbol{\mathcal{z}}^+|n,X,P,\mathscr{b}\rangle = \sqrt{n+1}|n+1,X,P,\mathscr{b}\rangle \end{cases} \tag{2.8}$$

so we can deduce

$$\begin{cases} \boldsymbol{\mathcal{p}}|n,X,P,\mathscr{b}\rangle = \dfrac{1}{\sqrt{2}}[\sqrt{n}|n-1,X,P,\mathscr{b}\rangle + \sqrt{n+1}|n+1,X,P,\mathscr{b}\rangle] \\ \boldsymbol{\mathcal{x}}|n,X,P,\mathscr{b}\rangle = \dfrac{i}{\sqrt{2}}[\sqrt{n}|n-1,X,P,\mathscr{b}\rangle - \sqrt{n+1}|n+1,X,P,\mathscr{b}\rangle] \end{cases} \tag{2.9}$$

$$\langle n,X,P,\mathscr{b}|\boldsymbol{\mathscr{p}}|\psi\rangle = \widehat{\widetilde{\boldsymbol{\mathscr{p}}}}\,\Psi^n(X,P,\mathscr{b})$$
$$= \frac{1}{\sqrt{2}}[\sqrt{n}\langle n-1,X,P,\mathscr{b}|\psi\rangle + \sqrt{n+1}\langle n+1,X,P,\mathscr{b}|\psi\rangle]$$
$$= \frac{1}{\sqrt{2}}\left[\sqrt{n}\Psi^{n-1}(X,P,\mathscr{b}) + \sqrt{n+1}\Psi^{n+1}(X,P,\mathscr{b})\right] \qquad (2.10)$$
$$\langle n,X,P,\mathscr{b}|\boldsymbol{\mathscr{x}}|\psi\rangle = \widehat{\widetilde{\boldsymbol{\mathscr{x}}}}\,\Psi^n(X,P,\mathscr{b})$$
$$= \frac{-i}{\sqrt{2}}[\sqrt{n}\langle n-1,X,P,\mathscr{b}|\psi\rangle - \sqrt{n+1}\langle n+1,X,P,\mathscr{b}|\psi\rangle]$$
$$= \frac{-i}{\sqrt{2}}\left[\sqrt{n}\Psi^{n-1}(X,P,\mathscr{b}) - \sqrt{n+1}\Psi^{n+1}(X,P,\mathscr{b})\right] \qquad (2.11)$$

From the expression of $\Psi^n$ [1],[2]

$$\Psi^n(X,P,\mathscr{b}) = \langle n,X,P,\mathscr{b}|\psi\rangle = \int \langle n,X,P,\mathscr{b}|x\rangle\,\langle x|\psi\rangle dx$$
$$= \int \varphi_n^*(x,X,P,\mathscr{b})\,\psi(x)dx$$
$$= \int \frac{H_n(\frac{x-X}{\sqrt{2}a})}{\sqrt{2^n n!\sqrt{2\pi}a}} e^{-\left(\frac{x-X}{2a}\right)^2 - iPx}\,\psi(x)dx \qquad (2.12)$$

we can deduce

$$\langle n,X,P,\mathscr{b}|\boldsymbol{\mathscr{p}}|\psi\rangle = \widehat{\widetilde{\boldsymbol{\mathscr{p}}}}\,\Psi^n(X,P,\mathscr{b})$$
$$= \frac{1}{\sqrt{2}}[\sqrt{n}\Psi^{n-1}(X,P,\mathscr{b}) + \sqrt{n+1}\Psi^{n+1}(X,P,\mathscr{b})]$$
$$= [\frac{\mathscr{b}}{\hbar}(i\hbar\frac{\partial}{\partial P} - X)]\Psi^n(X,P,\mathscr{b}) \qquad (2.13)$$
$$\langle n,X,P,\mathscr{b}|\boldsymbol{\mathscr{x}}|\psi\rangle = \widehat{\widetilde{\boldsymbol{\mathscr{x}}}}\,\Psi^n(X,P,\mathscr{b})$$
$$= \frac{i}{\sqrt{2}}[\sqrt{n}\langle n-1,X,P,\mathscr{b}|\psi\rangle - \sqrt{n+1}\langle n+1,X,P,\mathscr{b}|\psi\rangle]$$
$$= [\frac{a}{\hbar}(-i\hbar\frac{\partial}{\partial X} - P))]\Psi^n(X,P,\mathscr{b}) \qquad (2.14)$$

by identification we obtain

$$\begin{cases} \widehat{\widetilde{\boldsymbol{\mathscr{p}}}} = \dfrac{\mathscr{b}}{\hbar}(i\hbar\dfrac{\partial}{\partial P} - X) \\ \widehat{\widetilde{\boldsymbol{\mathscr{x}}}} = \dfrac{a}{\hbar}(-i\hbar\dfrac{\partial}{\partial X} - P) \end{cases} \qquad (2.15)$$

we verify that the commutation relation for $\widehat{\widetilde{\boldsymbol{\mathscr{x}}}}$ and $\widehat{\widetilde{\boldsymbol{\mathscr{p}}}}$ is

$$[\widehat{\widetilde{\boldsymbol{\mathscr{x}}}},\widehat{\widetilde{\boldsymbol{\mathscr{p}}}}]_- = i \qquad (2.16)$$

we also remark that more general expressions of $\widehat{\widetilde{\boldsymbol{\mathscr{x}}}}$ and $\widehat{\widetilde{\boldsymbol{\mathscr{p}}}}$ wich satisfy this commutation relation is

$$\begin{cases} \widehat{\widetilde{\boldsymbol{\mathscr{p}}}} = \dfrac{\mathscr{b}}{\hbar}(i\hbar\dfrac{\partial}{\partial P} - X) + \beta\dfrac{\partial}{\partial X} \\ \widehat{\widetilde{\boldsymbol{\mathscr{x}}}} = \dfrac{a}{\hbar}(-i\hbar\dfrac{\partial}{\partial X} - P) + \alpha\dfrac{\partial}{\partial P} \end{cases} \qquad (2.17)$$

in which $\beta$ and $\alpha$ are constant numbers.

Taking into account the relations

$$\begin{cases} \widehat{\widetilde{\boldsymbol{\mathscr{p}}}} = \dfrac{\widehat{\widetilde{\boldsymbol{p}}} - P}{\sqrt{2}\mathscr{b}} \\ \widehat{\widetilde{\boldsymbol{\mathscr{x}}}} = \dfrac{\widehat{\widetilde{\boldsymbol{x}}} - X}{\sqrt{2}a} \end{cases} \Leftrightarrow \begin{cases} \widehat{\widetilde{\boldsymbol{p}}} = \sqrt{2}\mathscr{b}\,\widehat{\widetilde{\boldsymbol{\mathscr{p}}}} + P \\ \widehat{\widetilde{\boldsymbol{x}}} = \sqrt{2}a\,\widehat{\widetilde{\boldsymbol{\mathscr{x}}}} + X \end{cases} \qquad (2.18)$$

we may deduce the expressions of $\widehat{\widetilde{\boldsymbol{p}}}$ and $\widehat{\widetilde{\boldsymbol{x}}}$ which satisfy the commutation relation $[\widehat{\widetilde{\boldsymbol{x}}},\widehat{\widetilde{\boldsymbol{p}}}]_- = i\hbar$ :

$$\begin{cases} \widehat{\widetilde{\boldsymbol{p}}} = \sqrt{2}\dfrac{\mathcal{B}}{\hbar}(i\hbar\dfrac{\partial}{\partial P} - X) + \sqrt{2}\mathscr{b}\beta\dfrac{\partial}{\partial X} + P \\ \widehat{\widetilde{\boldsymbol{x}}} = \sqrt{2}\dfrac{\mathcal{A}}{\hbar}(-i\hbar\dfrac{\partial}{\partial X} - P) + \sqrt{2}a\alpha\dfrac{\partial}{\partial P} + X \end{cases} \qquad (2.19)$$

in which, as in [2] , $\mathcal{B} = (\mathscr{b})^2$ and $\mathcal{A} = (a)^2$ and $\beta = \frac{a}{\mathscr{b}}\alpha$. In particular for the case $\beta = 0$ and $\alpha = 0$, we obtain the evident relations

$$\begin{cases} \widehat{\widetilde{p}} = \sqrt{2}\dfrac{\mathcal{B}}{\hbar}(i\hbar\dfrac{\partial}{\partial P} - X) + P \\ \widehat{\widetilde{x}} = \sqrt{2}\dfrac{\mathcal{A}}{\hbar}(-i\hbar\dfrac{\partial}{\partial X} - P) + X \end{cases} \qquad (2.20)$$

## III. MATRICES REPRESENTATIONS OF THE COORDINATE AND MOMENTUM OPERATORS

In the first section, we have established that for an observable $\boldsymbol{A}$, we can also have a matrix representation $[\boldsymbol{A}]$ in the basis $\{|n,X,P,\mathscr{b}\rangle\}$. The expression of the element $A_{\boldsymbol{m}}^{\boldsymbol{n}}$ of $[\boldsymbol{A}]$ is given by

$$A_{\boldsymbol{m}}^{\boldsymbol{n}} = \langle n,X,P,\mathscr{b}|\boldsymbol{A}|m,X,P,\mathscr{b}\rangle \qquad (3.1)$$

We have obtained

$$\begin{cases} \boldsymbol{\mathscr{p}}|n,X,P,\mathscr{b}\rangle = \dfrac{1}{\sqrt{2}}[\sqrt{n}|n-1,X,P,\mathscr{b}\rangle + \sqrt{n+1}|n+1,X,P,\mathscr{b}\rangle] \\ \boldsymbol{\mathscr{x}}|n,X,P,\mathscr{b}\rangle = \dfrac{i}{\sqrt{2}}[\sqrt{n}|n-1,X,P,\mathscr{b}\rangle - \sqrt{n+1}|n+1,X,P,\mathscr{b}\rangle] \end{cases} \qquad (3.2)$$

For the matrix representations $[\boldsymbol{\mathscr{p}}]$ and $[\boldsymbol{\mathscr{x}}]$ of $\boldsymbol{\mathscr{p}}$ and $\boldsymbol{\mathscr{x}}$ , the elements $\mathscr{p}_{\boldsymbol{m}}^{\boldsymbol{n}}$ and $\mathscr{x}_{\boldsymbol{m}}^{\boldsymbol{n}}$ are

$$\mathscr{p}_{\boldsymbol{m}}^{\boldsymbol{n}} = \langle n,X,P,\mathscr{b}|\boldsymbol{\mathscr{p}}|m,X,P,\mathscr{b}\rangle$$
$$= \frac{1}{\sqrt{2}}\left(\sqrt{m}\delta_{m-1}^n + \sqrt{m+1}\delta_{m+1}^n\right) \qquad (3.3)$$
$$\mathscr{x}_{\boldsymbol{m}}^{\boldsymbol{n}} = \langle n,X,P,\mathscr{b}|\boldsymbol{\mathscr{x}}|m,X,P,\mathscr{b}\rangle$$
$$= \frac{i}{\sqrt{2}}\left(\sqrt{m}\delta_{m-1}^n - \sqrt{m+1}\delta_{m+1}^n\right) \qquad (3.4)$$

Taking into account the relations

$$\boldsymbol{\mathscr{x}} = \frac{\boldsymbol{x} - X}{\sqrt{2}\mathscr{a}} \Rightarrow \boldsymbol{x} = \sqrt{2}\mathscr{a}\boldsymbol{\mathscr{x}} + X$$

$$\boldsymbol{\mathscr{p}} = \frac{\boldsymbol{p} - P}{\sqrt{2}\mathscr{b}} \Rightarrow \boldsymbol{p} = \sqrt{2}\mathscr{b}\boldsymbol{\mathscr{p}} + P$$

we can deduce for the elements $p_{\boldsymbol{m}}^{\boldsymbol{n}}$and $x_{\boldsymbol{m}}^{\boldsymbol{n}}$of thematrices representations $[\boldsymbol{p}]$ and $[\boldsymbol{x}]$ of $\boldsymbol{p}$ and $\boldsymbol{x}$

$$\begin{aligned} p_{\boldsymbol{m}}^{\boldsymbol{n}} &= \sqrt{2}\mathscr{b}\boldsymbol{\mathscr{p}}_{\boldsymbol{m}}^{\boldsymbol{n}} + P\delta_m^n \\ &= \mathscr{b}\left(\sqrt{m}\delta_{m-1}^n + \sqrt{m+1}\delta_{m+1}^n\right) + P\delta_m^n \end{aligned} \tag{3.5}$$

$$\begin{aligned} x_{\boldsymbol{m}}^{\boldsymbol{n}} &= \sqrt{2}\mathscr{a}\boldsymbol{\mathscr{x}}_{\boldsymbol{m}}^{\boldsymbol{n}} + X\delta_m^n \\ &= i\mathscr{a}\left(\sqrt{m}\delta_{m-1}^n - \sqrt{m+1}\delta_{m+1}^n\right) + X\delta_m^n \end{aligned} \tag{3.6}$$

We may verify that we have the relation

$$x_{\boldsymbol{n}}^{\boldsymbol{l}} p_{\boldsymbol{m}}^{\boldsymbol{n}} - p_{\boldsymbol{n}}^{\boldsymbol{l}} x_{\boldsymbol{m}}^{\boldsymbol{n}} = i\hbar\delta_m^l \tag{3.7}$$

## IV. MULTIDIMENSIONAL GENERALIZATION

As in our previous works [1], [2], [6], we use in this section notations based on [4].

For the multidimensional case, the commutation relation is

$$[\boldsymbol{\mathscr{p}}_\mu, \boldsymbol{\mathscr{x}}_\nu]_- = i\hbar\eta_{\mu\nu} \tag{4.1}$$

where $\eta_{\mu\nu}$ are the components of the metric tensor on Minkowski space

$$\eta_{\mu\nu} = \begin{cases} 1 \ if \ \mu = \nu = 0 \\ -1 \ if \ \mu = \nu = 1,2,3 \\ 0 \ \ if \ \mu \neq \nu \end{cases}$$

As in [2], we introduce the operators

$$\begin{cases} \boldsymbol{\mathscr{p}}_{\boldsymbol{\mu}} = \dfrac{\sqrt{2}}{\hbar}\mathscr{a}_\mu^\rho(\boldsymbol{p}_{\boldsymbol{\rho}} - P_\rho) \\ \boldsymbol{\mathscr{x}}_{\boldsymbol{\mu}} = \dfrac{\sqrt{2}}{\hbar}\mathscr{b}_\mu^\rho(\boldsymbol{x}_{\boldsymbol{\rho}} - X_\rho) \end{cases} \Leftrightarrow \begin{cases} \boldsymbol{p}_{\boldsymbol{\mu}} = \sqrt{2}\mathscr{b}_\mu^\rho\boldsymbol{\mathscr{p}}_{\boldsymbol{\rho}} + P_\rho \\ \boldsymbol{x}_{\boldsymbol{\mu}} = \sqrt{2}\mathscr{a}_\mu^\rho\boldsymbol{\mathscr{x}}_{\boldsymbol{\rho}} + X_\rho \end{cases} \tag{4.2}$$

in which $\mathscr{a}_\nu^\lambda$ and $\mathscr{b}_\mu^\rho$ are the multidimensional generalization of $\mathscr{a} = \Delta x$ and $\mathscr{b} = \Delta p$ as defined in [2]. They satisfy the relation

$$\mathscr{a}_\rho^\nu\mathscr{b}_\mu^\rho = \frac{\hbar}{2}\delta_\mu^\nu \tag{4.3}$$

where $\delta_\mu^\nu$ is the usual Kronecker 's symbol

$$\delta_\mu^\nu = \begin{cases} 1 \ if \ \mu = \nu \\ 0 \ \ if \mu \neq \nu \end{cases}$$

The commutation relation for the phase space representation $\tilde{\boldsymbol{\mathscr{p}}}_\mu$ and $\tilde{\boldsymbol{\mathscr{x}}}_\nu$of $\boldsymbol{\mathscr{p}}_{\boldsymbol{\mu}}$ and $\boldsymbol{\mathscr{x}}_{\boldsymbol{\nu}}$ is

$$[\tilde{\boldsymbol{\mathscr{p}}}_\mu, \tilde{\boldsymbol{\mathscr{x}}}_\nu]_- = i\eta_{\mu\nu} \tag{4.4}$$

It may be verified that general expressions satisfying this relation are

$$\begin{cases} \tilde{\boldsymbol{\mathscr{p}}}_\mu = \dfrac{\mathscr{b}_\mu^\rho}{\hbar}(-i\hbar\dfrac{\partial}{\partial P^\rho} - X_\rho) + \beta_\mu^\rho\dfrac{\partial}{\partial X^\rho} \\ \tilde{\boldsymbol{\mathscr{x}}}_\nu = \dfrac{\mathscr{a}_\nu^\lambda}{\hbar}(i\hbar\dfrac{\partial}{\partial X^\lambda} - P_\lambda) + \alpha_\nu^\lambda\dfrac{\partial}{\partial P^\lambda} \end{cases} \tag{4.5}$$

in which $\beta_\mu^\rho$ and $\alpha_\nu^\lambda$ are constant numbers. It may be chosen in particular $\beta_\mu^\rho = 0$ and $\alpha_\mu^\rho = 0$, in that case

$$\begin{cases} \tilde{\boldsymbol{\mathscr{p}}}_\mu = \dfrac{\mathscr{b}_\mu^\rho}{\hbar}(-i\hbar\dfrac{\partial}{\partial P^\rho} - X_\rho) \\ \tilde{\boldsymbol{\mathscr{x}}}_\nu = \dfrac{\mathscr{a}_\nu^\lambda}{\hbar}(i\hbar\dfrac{\partial}{\partial X^\lambda} - P_\lambda) \end{cases} \tag{4.6}$$

And taking into account (4.2), we obtain for $\tilde{\boldsymbol{p}}_\mu$and $\tilde{\boldsymbol{x}}_\nu$ :

$$\begin{cases} \tilde{\boldsymbol{p}}_{\boldsymbol{\mu}} = \sqrt{2}\dfrac{\mathcal{B}_\mu^\sigma}{\hbar}(-i\hbar\dfrac{\partial}{\partial P^\sigma} - X_\sigma) + P_\mu \\ \tilde{\boldsymbol{x}}_{\boldsymbol{\mu}} = \sqrt{2}\dfrac{\mathcal{A}_\mu^\lambda}{\hbar}(i\hbar\dfrac{\partial}{\partial X^\lambda} - P_\lambda) + X_\mu \end{cases} \tag{4.7}$$

in which, as in [4], $\mathcal{B}_\mu^\sigma = \mathscr{b}_\mu^\rho\mathscr{b}_\rho^\sigma$ and $\mathcal{A}_\mu^\lambda = \mathscr{a}_\mu^\rho\mathscr{a}_\rho^\lambda$.

## V. DISPERSION OPERATORS

In our previous works [2], we have introduced the generators $\beth_{\boldsymbol{\mu\nu}}^+$,$\beth_{\boldsymbol{\mu\nu}}^-$and $\beth_{\boldsymbol{\mu\nu}}^\times$of the dispersion operators algebra:

$$\begin{cases} \beth_{\boldsymbol{\mu\nu}}^+ = 4\mathscr{b}_\mu^\rho\mathscr{b}_\nu^\lambda\bar{\beth}_{\boldsymbol{\rho\lambda}}^+ \\ \beth_{\boldsymbol{\mu\nu}}^- = 4\mathscr{b}_\mu^\rho\mathscr{b}_\nu^\lambda\bar{\beth}_{\boldsymbol{\rho\lambda}}^- \\ \beth_{\boldsymbol{\mu\nu}}^\times = 4\boldsymbol{\mathscr{b}}_\mu^\rho\mathscr{b}_\nu^\lambda\bar{\beth}_{\boldsymbol{\rho\lambda}}^\times \end{cases} \tag{5.1}$$

in which

$$\begin{cases} \bar{\beth}_{\boldsymbol{\mu\nu}}^+ = \dfrac{1}{4}(\boldsymbol{\mathscr{p}}_\mu\boldsymbol{\mathscr{p}}_\nu + \boldsymbol{\mathscr{x}}_\mu\boldsymbol{\mathscr{x}}_\nu) \\ \bar{\beth}_{\boldsymbol{\mu\nu}}^- = \dfrac{1}{4}(\boldsymbol{\mathscr{p}}_\mu\boldsymbol{\mathscr{p}}_\nu - \boldsymbol{\mathscr{x}}_\mu\boldsymbol{\mathscr{x}}_\nu) \\ \bar{\beth}_{\boldsymbol{\mu\nu}}^\times = \dfrac{1}{4}(\boldsymbol{\mathscr{p}}_\mu\boldsymbol{\mathscr{x}}_\nu + \boldsymbol{\mathscr{x}}_\nu\boldsymbol{\mathscr{p}}_\mu) \end{cases} \tag{5.2}$$

Then, we have in the phase space representation

$$\begin{cases} \tilde{\bar{\beth}}_{\boldsymbol{\mu\nu}}^+ = \dfrac{1}{4}(\tilde{\boldsymbol{\mathscr{p}}}_\mu\tilde{\boldsymbol{\mathscr{p}}}_\nu + \tilde{\boldsymbol{\mathscr{x}}}_\mu\tilde{\boldsymbol{\mathscr{x}}}_\nu) \\ \tilde{\bar{\beth}}_{\boldsymbol{\mu\nu}}^- = \dfrac{1}{4}(\tilde{\boldsymbol{\mathscr{p}}}_\mu\tilde{\boldsymbol{\mathscr{p}}}_\nu - \tilde{\boldsymbol{\mathscr{x}}}_\mu\tilde{\boldsymbol{\mathscr{x}}}_\nu) \\ \tilde{\bar{\beth}}_{\boldsymbol{\mu\nu}}^\times = \dfrac{1}{4}(\tilde{\boldsymbol{\mathscr{p}}}_\mu\tilde{\boldsymbol{\mathscr{x}}}_\nu + \tilde{\boldsymbol{\mathscr{x}}}_\nu\tilde{\boldsymbol{\mathscr{p}}}_\mu) \end{cases} \tag{5.3}$$

If we use the expression (4.5) of $\tilde{\boldsymbol{\mathscr{p}}}_\mu$ and $\tilde{\boldsymbol{\mathscr{x}}}_\nu$, we obtain

$$\tilde{\boldsymbol{\mathscr{p}}}_\mu\tilde{\boldsymbol{\mathscr{p}}}_\nu = \left[\frac{\mathscr{b}_\mu^\rho}{\hbar}(-i\hbar\frac{\partial}{\partial P^\rho} - X_\rho)\right]\left[\frac{\mathscr{b}_\nu^\sigma}{\hbar}(-i\hbar\frac{\partial}{\partial P^\sigma} - X_\sigma)\right]$$

$$= \frac{\mathscr{b}_\mu^\rho\mathscr{b}_\nu^\sigma}{\hbar^2}[-\hbar^2\frac{\partial}{\partial P^\rho}\frac{\partial}{\partial P^\sigma} + i\hbar(X_\sigma\frac{\partial}{\partial P^\rho} + X_\rho\frac{\partial}{\partial P^\sigma}) + X_\rho X_\sigma]$$

$$\hat{\boldsymbol{x}}_{\mu}\hat{\boldsymbol{x}}_{\nu}=\left[\frac{a_{\mu}^{\lambda}}{\hbar}(i\hbar\frac{\partial}{\partial X^{\lambda}}-P_{\lambda})\right]\left[\frac{a_{\nu}^{\rho}}{\hbar}(i\hbar\frac{\partial}{\partial X^{\rho}}-P_{\rho})\right]$$

$$=\frac{a_{\mu}^{\lambda}a_{\nu}^{\rho}}{\hbar^{2}}[-\hbar^{2}\frac{\partial}{\partial X^{\lambda}}\frac{\partial}{\partial X^{\rho}}-i\hbar(P_{\rho}\frac{\partial}{\partial X^{\lambda}}+P_{\lambda}\frac{\partial}{\partial X^{\rho}})+P_{\lambda}P_{\rho}]$$

$$\hat{\boldsymbol{p}}_{\mu}\hat{\boldsymbol{x}}_{\nu}=\left[\frac{b_{\mu}^{\rho}}{\hbar}(-i\hbar\frac{\partial}{\partial P^{\rho}}-X_{\rho})\right]\left[\frac{a_{\nu}^{\lambda}}{\hbar}(i\hbar\frac{\partial}{\partial X^{\lambda}}-P_{\lambda})\right]=$$

$$\frac{b_{\mu}^{\rho}a_{\nu}^{\lambda}}{\hbar^{2}}[\hbar^{2}\frac{\partial}{\partial P^{\rho}}\frac{\partial}{\partial X^{\lambda}}+i\hbar(\delta_{\rho\lambda}+P_{\lambda}\frac{\partial}{\partial P^{\rho}}-X_{\rho}\frac{\partial}{\partial X^{\lambda}})+X_{\rho}P_{\lambda}]$$

$$\hat{\boldsymbol{x}}_{\nu}\hat{\boldsymbol{p}}_{\mu}=\left[\frac{a_{\nu}^{\lambda}}{\hbar}(i\hbar\frac{\partial}{\partial X^{\lambda}}-P_{\lambda})\right]\left[\frac{b_{\mu}^{\rho}}{\hbar}(-i\hbar\frac{\partial}{\partial P^{\rho}}-X_{\rho})\right]=$$

$$\frac{b_{\mu}^{\rho}a_{\nu}^{\lambda}}{\hbar^{2}}[\hbar^{2}\frac{\partial}{\partial X^{\lambda}}\frac{\partial}{\partial P^{\rho}}-i\hbar(\delta_{\rho\lambda}+X_{\rho}\frac{\partial}{\partial X^{\lambda}}-P_{\lambda}\frac{\partial}{\partial P^{\rho}})+P_{\lambda}X_{\rho}]$$

So we have

$$\hat{\mathfrak{I}}_{\mu\nu}^{+}=\frac{1}{4}(\hat{\boldsymbol{p}}_{\mu}\hat{\boldsymbol{p}}_{\nu}+\hat{\boldsymbol{x}}_{\mu}\hat{\boldsymbol{x}}_{\nu})=$$

$$\frac{1}{4}\{\frac{b_{\mu}^{\rho}b_{\nu}^{\lambda}}{\hbar^{2}}[-\hbar^{2}\frac{\partial^{2}}{\partial P^{\rho}\partial P^{\lambda}}+i\hbar(X_{\lambda}\frac{\partial}{\partial P^{\rho}}+X_{\rho}\frac{\partial}{\partial P^{\lambda}})+X_{\rho}X_{\lambda}]$$

$$+\frac{a_{\mu}^{\rho}a_{\nu}^{\lambda}}{\hbar^{2}}[-\hbar^{2}\frac{\partial^{2}}{\partial X^{\rho}\partial X^{\lambda}}-i\hbar(P_{\lambda}\frac{\partial}{\partial X^{\rho}}+P_{\rho}\frac{\partial}{\partial X^{\lambda}})+P_{\rho}P_{\lambda}]\}$$

$$\hat{\mathfrak{I}}_{\mu\nu}^{-}=\frac{1}{4}(\hat{\boldsymbol{p}}_{\mu}\hat{\boldsymbol{p}}_{\nu}-\hat{\boldsymbol{x}}_{\mu}\hat{\boldsymbol{x}}_{\nu})=$$

$$\frac{1}{4}\{\frac{b_{\mu}^{\rho}b_{\nu}^{\lambda}}{\hbar^{2}}[-\hbar^{2}\frac{\partial^{2}}{\partial P^{\rho}\partial P^{\lambda}}+i\hbar(X_{\lambda}\frac{\partial}{\partial P^{\rho}}+X_{\rho}\frac{\partial}{\partial P^{\lambda}})+X_{\rho}X_{\lambda}]$$

$$-\frac{a_{\mu}^{\rho}a_{\nu}^{\lambda}}{\hbar^{2}}[-\hbar^{2}\frac{\partial^{2}}{\partial X^{\rho}\partial X^{\lambda}}-i\hbar(P_{\lambda}\frac{\partial}{\partial X^{\rho}}+P_{\rho}\frac{\partial}{\partial X^{\lambda}})+P_{\rho}P_{\lambda}]\}$$

$$\hat{\mathfrak{I}}_{\mu\nu}^{\times}=\frac{1}{4}(\hat{\boldsymbol{p}}_{\mu}\hat{\boldsymbol{x}}_{\nu}+\hat{\boldsymbol{x}}_{\nu}\hat{\boldsymbol{p}}_{\mu})=$$

$$\frac{1}{2}\frac{b_{\mu}^{\rho}a_{\nu}^{\lambda}}{\hbar^{2}}[\hbar^{2}\frac{\partial^{2}}{\partial P^{\rho}\partial X^{\lambda}}+i\hbar(P_{\lambda}\frac{\partial}{\partial P^{\rho}}-X_{\rho}\frac{\partial}{\partial X^{\lambda}})+P_{\lambda}X_{\rho}]$$

Then, using the relation (5.1), we obtain

$$\hat{\mathfrak{I}}_{\mu\nu}^{+}=4b_{\mu}^{\varepsilon}b_{\nu}^{\theta}\hat{\mathfrak{I}}_{\varepsilon\theta}^{+}$$

$$=\frac{\mathcal{B}_{\mu}^{\rho}\mathcal{B}_{\nu}^{\lambda}}{\hbar^{2}}[-\hbar^{2}\frac{\partial}{\partial P^{\rho}}\frac{\partial}{\partial P^{\lambda}}+i\hbar(X_{\lambda}\frac{\partial}{\partial P^{\rho}}+X_{\rho}\frac{\partial}{\partial P^{\lambda}})+X_{\rho}X_{\lambda}]$$

$$+\frac{1}{4}[-\hbar^{2}\frac{\partial}{\partial X^{\mu}}\frac{\partial}{\partial X^{\nu}}+i\hbar(P_{\mu}\frac{\partial}{\partial X^{\nu}}+P_{\nu}\frac{\partial}{\partial X^{\mu}})+P_{\mu}P_{\nu}]$$

$$\hat{\mathfrak{I}}_{\mu\nu}^{-}=4b_{\mu}^{\varepsilon}b_{\nu}^{\theta}\hat{\mathfrak{I}}_{\varepsilon\theta}^{-}$$

$$=\frac{\mathcal{B}_{\mu}^{\rho}\mathcal{B}_{\nu}^{\lambda}}{\hbar^{2}}[-\hbar^{2}\frac{\partial}{\partial P^{\rho}}\frac{\partial}{\partial P^{\lambda}}+i\hbar(X_{\lambda}\frac{\partial}{\partial P^{\rho}}+X_{\rho}\frac{\partial}{\partial P^{\lambda}})+X_{\rho}X_{\lambda}]$$

$$-\frac{1}{4}[-\hbar^{2}\frac{\partial}{\partial X^{\mu}}\frac{\partial}{\partial X^{\nu}}+i\hbar(P_{\mu}\frac{\partial}{\partial X^{\nu}}+P_{\nu}\frac{\partial}{\partial X^{\mu}})+P_{\mu}P_{\nu}]$$

$$\hat{\mathfrak{I}}_{\mu\nu}^{\times}=4b_{\mu}^{\varepsilon}b_{\nu}^{\theta}\hat{\mathfrak{I}}_{\varepsilon\theta}^{\times}$$

$$=\frac{\mathcal{B}_{\mu}^{\rho}}{\hbar}[\hbar^{2}\frac{\partial^{2}}{\partial P^{\rho}\partial X^{\nu}}+i\hbar(P_{\nu}\frac{\partial}{\partial P^{\rho}}-X_{\rho}\frac{\partial}{\partial X^{\nu}})+P_{\nu}X_{\rho}]$$

## VI. CONCLUSION

From our previous works [1], [2] we have shown in this paper that there are two possibilities for the representation of coordinate and momentum operators in the framework of the phase space representation. The first is a representation with differential operators and the second one is matrices representation.
These possibilities are linked with the fact that there are two ways to expand a state in the basis $\{|n,X,P,\Delta p\rangle\}$ as given in the relations (1.18) and (1.21).

The establishment of the expressions of the differential operator representations are performed in the section 2 and the matrices one in the section 3. The results are presented in (2.19), (3.5) and (3.6).
Multidimensional generalization of the differential operators representation of coordinate and momentum are established in the section 4 and the representations of dispersion operator are given in section 5.

As expected, the main results in this works are the establishment of the new representation of coordinate, momentum and dispersion operators which correspond to the phase space representation.